\documentclass[10pt,prd,superscriptaddress,preprintnumbers,amsmath,amssymb,nofootinbib]{revtex4}
\usepackage{epsfig}  
\usepackage{graphicx}
\usepackage{hyperref}
\usepackage{color}
\usepackage{float}
\usepackage{amsfonts}
\usepackage{amsmath}
\usepackage{slashed}
\usepackage{soul}

\large

\begin{document}

\title{Solutions to $R_D$-$R_{D^*}$ in light of Belle 2019 data}

\author{Ashutosh Kumar Alok}
\email{akalok@iitj.ac.in}
\affiliation{Indian Institute of Technology Jodhpur, Jodhpur 342037, India}

\author{Dinesh Kumar}
\email{dinesh.kumar@ncbj.gov.pl}
\affiliation{National Centre for Nuclear Research, Warsaw, Poland}
\affiliation{Department of Physics, University of Rajasthan, Jaipur 302004, India}

\author{Suman Kumbhakar}
\email{suman@phy.iitb.ac.in}
\affiliation{Indian Institute of Technology Bombay, Mumbai 400076, India}

\author{S Uma Sankar}
\email{uma@phy.iitb.ac.in}
\affiliation{Indian Institute of Technology Bombay, Mumbai 400076, India}

\begin{abstract}
Earlier this year, the Belle collaboration presented their new measurements of $R_D$ and $R_{D^*}$ using a new method. These measurements are consistent with the Standard Model predictions, whereas the global averages of the earlier measurements had a $4.1\sigma$ discrepancy. With the inclusion of the new data in the global averages, the discrepancy comes down to $3.1\sigma$. In this work, we study the study the new physics solutions to the $R_D$-$R_{D^*}$ anomaly allowed by the reduction in the discrepancy. Among the four fermion operators, which arise through a single particle exchange, only the $(V-A)$ operator solution survives. We found three additional solutions with two dis-similar operators. The branching ratio of $B_c\rightarrow \tau\,\bar{\nu}$ is powerful discriminant between these four allowed solutions.


\end{abstract}
 
\maketitle 

\section{Introduction} 
The flavor ratios $R_{D^{(*)}} = \Gamma(B\rightarrow D^{(*)}\,\tau\,\bar{\nu})/ \Gamma(B\rightarrow D^{(*)}\, \{e/\mu\} \, \bar{\nu})$ were measured by BaBar~\cite{Lees:2012xj,Lees:2013uzd}, Belle~\cite{Huschle:2015rga,Sato:2016svk,Hirose:2016wfn} and LHCb~\cite{Aaij:2015yra} collaborations. The average values of these measurements differ from their respective Standard Model (SM) predictions by 3.9$\sigma$~\cite{hflav-2016}. In all these measurements, the $\tau$ lepton was not reconstructed but was identified through other kinematical information. In ref.~\cite{Aaij:2017uff}, LHCb collaboration attempted to reconstruct the $\tau$ lepton through its $3\pi$ decay mode, in making a separate measurement of $R_{D^*}$. Post this measurement, the discrepancy of $R_D$-$R_{D^*}$ data with SM predictions increased to $4.1\sigma$~\cite{hflav-2017}. The observed values of $R_D$ and $R_{D^*}$ are noticeably higher than their respective SM predictions in all these measurements~\cite{Amhis:2016xyh}. 
These measurements indicate the violation of lepton flavor universality. The higher values of $R_D$ and $R_{D^*}$ are assumed to occur due to new physics (NP) contribution to the $b\rightarrow c\,\tau\,\bar{\nu}$ decay. New physics in $b\rightarrow c\, \{e/\mu\}\,\bar{\nu}$ is ruled out by other data~\cite{Alok:2017qsi}. LHCb collaboration also measured the related flavor ratio $R_{J/\psi} = \Gamma(B_c\rightarrow J/\psi \, \tau \, \bar{\nu})/\Gamma(B_c\rightarrow J/\psi \, \mu\, \bar{\nu})$ and found it to be $1.7\sigma$ higher than the SM prediction~\cite{Aaij:2017tyk}.

In the SM, the charged current transition $b\rightarrow c\,\tau\,\bar{\nu}$ occurs at tree level. To account for the measured higher values of flavor ratios, the NP amplitudes are expected to be about $10\%$ of the SM amplitude. The complete list of effective operators leading to $b\rightarrow c\,\tau\,\bar{\nu}$ decay are listed in ref.~\cite{Freytsis:2015qca}. These operators can be classified by their Lorentz structure. Different Lorentz structures contribute differently to the flavor ratios. The coefficients of these operators are determined by fitting the theoretical predictions to the data. The purely leptonic decay $B_c\rightarrow \tau\,\bar{\nu}$ is also driven by these operators. This decay mode has not been observed yet but the total decay width of $B_c$ meson has been measured. In the SM, the branching ratio for this mode is small because of helicity suppression. The constraint that $\Gamma(B_c\rightarrow \tau\,\bar{\nu})_{\rm NP}$ should be less than the measured decay width of $B_c$ meson leads to useful constraints on a class of NP operators. 

In addition to the branching ratios, it is possible to measure various other quantities in $B\rightarrow D^*\,\tau\bar{\nu}$ decay. The polarization fractions of the $\tau$ lepton ($P_{\tau}^{D^*}$)~\cite{Hirose:2016wfn} and the $D^*$ meson ($f_L^{D^*}$)~\cite{Alok:2016qyh} are two such quantities which can be measured even without the reconstruction of $\tau$ lepton. These observables can lead to discrimination between different NP operators. If the $\tau$ lepton is reconstructed and its momentum determined then it is possible to measure two more angular observables, the forward-backward asymmetry $A_{FB}^{D^*}$ and longitudinal-transverse asymmetry $A^{D^*}_{LT}$~\cite{Alok:2010zd}. If these asymmetries are measured then it can lead to further discrimination between NP operators~\cite{Alok:2018uft}.

The new physics can be parametrized in terms of five different operators $\mathcal{O}_i$, with different Lorentz structures. They are
\begin{equation}
\mathcal{O}_{V_L} =(\bar c \gamma_{\mu} P_L b)(\bar \tau \gamma^{\mu}  P_L \nu) \ , \quad  
\mathcal{O}_{V_R}=(\bar c \gamma_{\mu}  P_R b)(\bar \tau \gamma^{\mu}  P_L \nu) \nonumber\ , \quad
\end{equation}  
\begin{equation}
\mathcal{O}_{S_L}=(\bar c P_L b)(\bar \tau P_L \nu ), \quad
\mathcal{O}_{S_R}=(\bar c P_R b)(\bar \tau P_L \nu),  \quad    
\mathcal{O}_T=(\bar c \sigma_{\mu \nu}P_L b)(\bar \tau \sigma^{\mu \nu} P_L \nu) \ . 
\label{ops}
\end{equation}
In writing the above operators, we assumed that the neutrino is purely a left chiral fermion. These operators appear in the effective Hamiltonian with coefficients $\tilde{C}_i$, where we assume $\tilde{C}_i$ are real. First we consider the effect of each individual $\mathcal{O}_i$ on $R_D$-$R_{D^*}$ anomaly. 
\begin{itemize}
\item The operator $\mathcal{O}_{V_L}$ has the same Lorentz structure as the SM operator. This amplitude adds to SM amplitude and hence $R_D$ and $R_{D^*}$ become proportional to $(1+\tilde{C}_{V_L})^2$. A fit to data gives a solution for $\tilde{C}_{V_L}$ because the fractional increase in $R_D$ and $R_{D^*}$ are roughly the same.
\item If the NP operator is $\mathcal{O}_{V_R}$, $R_D$ is proportional to $(1+\tilde{C}_{V_R})^2$ where as $R_{D^*}$ depends to a large extent on $(1-\tilde{C}_{V_R})^2$. Given the data, it is not possible to find a common solution to both $R_D$ and $R_{D^*}$.~\footnote{NP in the form of only $O_{V_R}$ is allowed if $\tilde{C}_{V_R}$ is allowed to be complex~\cite{Iguro:2018vqb}.}
\item The operators $\mathcal{O}_{S_L}$ and $\mathcal{O}_{S_R}$ contain the pseudoscalar bilinear $\bar{c}\gamma_5b$. Hence the amplitudes due to these operators are not subject to helicity suppression. These amplitudes predict large branching ratios for $B_c\rightarrow \tau\,\bar{\nu}$. Therefore, the constraint on this branching ratio restricts the solutions given by $R_D$-$R_{D^*}$ fit.
\item The tensor operator $\mathcal{O}_T$ solution with large Wilson coefficient predicts $f_L^{D^*}$ to be  much smaller than the predicted values of other solutions~\cite{Alok:2016qyh}. Hence an accurate measurement of this polarization fraction can distinguish this solution from others.  
\end{itemize}

Last year, Belle collaboration announced the first measurement of $f_L^{D^*}$~\cite{Adamczyk:2019wyt,Abdesselam:2019wbt}. Earlier this year, Belle collaboration announced a new measurement of $R_D$ and $R_{D^*}$~\cite{Abdesselam:2019dgh}, which is consistent with the SM prediction. Inclusion of this measurement in computing a new world average brings down the discrepancy with SM from $4.1\sigma$ to $3.1\sigma$. This is still a substantial discrepancy. Moreover, the central values of the new measurement are also higher than the SM predictions. This has been the feature of all $R_D$ and $R_{D^*}$ measurements no matter what the discrepancy is. Given that the measured deviation from the SM prediction is always positive, it is expected that there is indeed new physics present. In this work, we study the effect of these two recent Belle measurements on the previously obtained solutions to $R_D$-$R_{D^*}$ anomaly~\cite{Alok:2017qsi}. We find that only the $\mathcal{O}_{V_L}$ solution survives among these.

\section{NP solutions arising through one particle exchange}

The most general four-fermion effective Hamiltonian for $b\rightarrow c\,\tau\,\bar{\nu}$ transition can be parametrized as \cite{Freytsis:2015qca}
\begin{eqnarray}
H_{eff} &=& \frac{4 G_F}{\sqrt{2}} V_{cb} \left[\mathcal{O}_{V_L} + \frac{\sqrt{2}}{4 G_F V_{cb}\Lambda^2}  \sum_i C^{(','')}_i\mathcal{O}^{(','')}_i\right]\,, \nonumber\\
&=& \frac{4 G_F}{\sqrt{2}} V_{cb} \left[\mathcal{O}_{V_L} + \alpha  \sum_i C^{(','')}_i\mathcal{O}^{(','')}_i\right]\,,
\label{effH}
\end{eqnarray}
where we defined $(2\sqrt{2} G_F V_{cb}\Lambda^2)^{-1} \equiv \alpha$.  We assume the new physics scale, $\Lambda$, to be 1 TeV which leads to $\alpha = 0.749$. The unprimed operators are defined in eq.~(\ref{ops}).
The primed operators couple
a bilinear of form $\bar{\tau} \Gamma b$ to the bilinear $\bar{c}\Gamma \nu$, whereas the 
double primed operators are products of the bilinears $\bar{\tau} \Gamma c^c$ and $\bar{b^c}\Gamma \nu$. 
Each of these primed and double primed operators can be expressed in terms of the corresponding
unprimed operators through Fierz transforms. These operators and their Fierz transformed forms are listed in ref. \cite{Freytsis:2015qca}.
 Within the SM, only $\mathcal{O}_{V_L}$ operator is present. The NP operators $\mathcal{O}_i$, $\mathcal{O}^{'}_i$ 
and $\mathcal{O}^{''}_i$ include all other possible Lorentz structures. The NP effects are encoded in the Wilson coefficients  $C_i, C^{'}_i$ and $C^{''}_i$, which we assume to be real.

In a previous work~\cite{Alok:2017qsi}, we did a $\chi^2$ fit to the data on $R_D$, $R_{D^*}$, $R_{J/\psi}$ and $P_{\tau}^{D^*}$, available up to the summer of 2017. We used the following data in this fit:
\begin{eqnarray}
R_D & = &  0.407 \pm 0.039 \pm 0.024,\nonumber \\
R_{D^*} & =& 0.304 \pm 0.013 \pm 0.007,\nonumber\\
R_{J/\psi} & =& 0.71\pm 0.17\pm 0.18,\nonumber\\
P_{\tau}^{D^*} &= & -0.38\pm 0.51^{+0.21}_{-0.16}.
\end{eqnarray}
The data of $R_D$ and $R_{D^*}$ are taken from ref.~\cite{hflav-2017}. That of $R_{J/\psi}$ and $P_{\tau}^{D^*}$ are taken from refs.~\cite{Aaij:2017tyk} and \cite{Hirose:2016wfn} respectively. In doing this fit, we have taken into account the correlation between the measured values of $R_D$ and $R_{D^*}$. The  $B\rightarrow D^{(*)}\, l\,  \bar{\nu}$ decay distributions depend upon hadronic form-factors. So far, the determination of these form-factors depends heavily on HQET techniques.  In this work we use the HQET form factors, parametrized by Caprini {\it et al.} \cite{Caprini:1997mu}. The parameters for $B\rightarrow D$ decay are well known in lattice QCD \cite{Aoki:2016frl} and we use them in our analyses. For $B\rightarrow D^*$ decay, the HQET parameters are extracted using data from Belle and BaBar experiments along with lattice inputs. In this work, the numerical values of these parameters are taken from refs. \cite{Bailey:2014tva} and  \cite{Amhis:2016xyh}.

The previous analysis was performed under two different assumptions: (i) only one NP operator is present and (ii) two similar NP operators are present. This was based on the assumption that these operators arise through the exchange of only one new particle. The allowed solutions satisfied the constraints (a) $\chi^2_{\rm min}\leq 4.8$ and (b) $\mathcal{B}(B_c\rightarrow \tau\,\bar{\nu})<10\%$~\cite{Akeroyd:2017mhr}. The strong constraint on $\mathcal{B}(B_c\rightarrow \tau\,\bar{\nu})$ is obtained from LEP upper limit on the effective branching ratio of charged $B$ mesons to $\tau\bar{\nu}$~\cite{Acciarri:1996bv}, where the ratio of production of $B_c$ to $B_u$ mesons is assumed to be $f_c/f_u$. The fraction $f_c/f_u$ is estimated from the data on $B_u$ and $B_c$ decays at Tevatron~\cite{Abe:1998fb, Abulencia:2006zu} and at LHCb~\cite{Aaij:2014jxa}. We obtained three solutions with the single operator assumption and three more with the two similar operators assumption. These solutions are listed in table~\ref{one}.  
\begin{table}[h!]
\centering
\tabcolsep 7pt
\begin{tabular}{|c|c|c|}
\hline\hline
NP type & Best fit value(s) & $\chi^2_{\rm min}$  \\
\hline
SM  & $C_{i}=0$ & 22.44  \\
\hline
$C_{V_L}$  &   $0.15 \pm 0.03$  & 2.9  \\
\hline
$C_T$  &  $0.52 \pm 0.02$ & 4.8   \\
\hline
$C''_{S_L}$ & $-0.53\pm 0.10$ & 2.9  \\
\hline
$(C_{V_L},C_{V_R})$& $(-1.29, 1.51)$& 2.1  \\
\hline
$(C'_{V_L},\, C'_{V_R})$  &  $(0.12, -0.06)$ &2.1  \\
\hline
$(C''_{S_L},\, C''_{S_R})$  & $(-0.64, -0.08)$  &2.0  \\
\hline\hline
\end{tabular}
\caption{ Best fit values of the coefficients of new physics operators at $\Lambda = 1$ TeV by making use of data of $R_D$, $R_{D^*}$, $R_{J/\psi}$ and $P_{\tau}^{D^*}$, taken from ref.~\cite{Alok:2017qsi}. In this fit, we use the updated summer 2017 world averages of $R_D$-$R_{D^*}$. Here we allow only those solutions for which $\chi^2_{\rm min}\leq 4.8$ as well as $\mathcal{B}(B_c\rightarrow \tau\,\bar{\nu})< 10\%$. }
\label{one}
\end{table}

In the first set, there is a tensor operator solution with the coefficient $C_T = 0.52$. This solution predicts the $D^*$ polarization fraction to be $f_L^{D^*} \approx 0.14\pm 0.04$~\cite{Alok:2016qyh}. The prediction for each of the other solutions is $f_L^{D^*} \approx 0.46\pm 0.04$, which is also the SM prediction.

During the past year, the Belle experiment announced two new results: 
\begin{itemize}
\item They made the first measurement of $f_L^{D^*}$. The measured value, $0.60\pm 0.08\,(\rm stat.)\pm 0.04\,(\rm syst.)$~\cite{Adamczyk:2019wyt,Abdesselam:2019wbt}, is about $1.5\sigma$  above the SM prediction but is $4.5\sigma$ away from the prediction of the $C_T = 0.52$ solution. Hence, this measurement completely rules out the tensor solution.
\item At Moriond 2019, they also presented new measurements: $R_D = 0.307\pm 0.037\pm 0.016$ and $R_{D^*} = 0.283\pm 0.018\pm 0.014$~\cite{Abdesselam:2019dgh}. These are consistent with the SM predictions:  $R_D|_{\rm SM} = 0.299\pm 0.003$ and $R_{D^*}|_{\rm SM} = 0.258\pm 0.005$~\cite{Amhis:2016xyh}.
\end{itemize}
Including these new measurements in the global averages leads to $R_D = 0.340\pm 0.027\pm 0.013$ and $R_{D^*} = 0.295\pm 0.011\pm 0.008$~\cite{avg19}. The discrepancy between these values and the SM predictions is down to $3.1\sigma$ from $4.1\sigma$. It should be noted that the central values of the new measurements also are higher than the SM predictions, which has been a common feature of all the $R_D$-$R_{D^*}$ measurements, as mentioned in the introduction. 

We take this consistent positive deviations to be an indication for the presence of new physics. We re-did our analysis with the new global averages for $R_D$ and $R_{D^*}$ along with $R_{J/\psi}$, $P_{\tau}^{D^*}$ and $f_L^{D^*}$. In this analysis, we included the renormalization group (RG) effects in the evolution of the WCs from the scale $\Lambda = 1$ TeV to the scale $m_b$~\cite{Gonzalez-Alonso:2017iyc}. These effects are particularly important for the scalar and tensor operators.

We select the NP solutions satisfying the constraints  $\chi^2_{\rm min}\leq 5$ as well as $\mathcal{B}(B_c\rightarrow \tau\bar{\nu})< 10\%$. We raised the upper limit on $\chi^2_{\rm min}$ because the re-fit included an extra data point on $f_L^{D^*}$. Among the solutions listed in table~\ref{one}, we note that only the $\mathcal{O}_{V_L}$ solution survives among the single operator solutions. However, its coefficient is reduced by a third to $C_{V_L} = 0.10\pm 0.02$. Among the two similar operator solutions, only the $(\mathcal{O}''_{S_L},\, \mathcal{O}''_{S_R})$ persists in principle, with the WCs $(C''_{S_L},\, C''_{S_R}) = (0.05, 0.24)$. The value of $C''_{S_L}$ is quite small, $C''_{S_R} \approx 2C_{V_L}$ and the Fierz transform of $\mathcal{O}''_{S_R}$ is $\mathcal{O}_{V_L}/2$. Therefore, this solution is effectively equivalent to the $\mathcal{O}_{V_L}$ solution. Among the single operator and two similar operator solutions, \textbf{only the $\mathcal{O}_{V_L}$ solution is allowed by the present data}.

 
\section{NP solutions with mixed spin operators}
 
As we saw in the previous section, the present data allow only the $\mathcal{O}_{V_L}$ solution, among the NP operators arising from a single particle exchange. To explore the full set of NP solutions, here we consider the possibility of two dis-similar NP operators being present in the new physics Hamiltonian. This additional possibility must be considered because a NP model is likely to contain a number of new particles of different spins.  

\begin{figure*}[h] 
\centering
\begin{tabular}{cc}
\includegraphics[width=65mm]{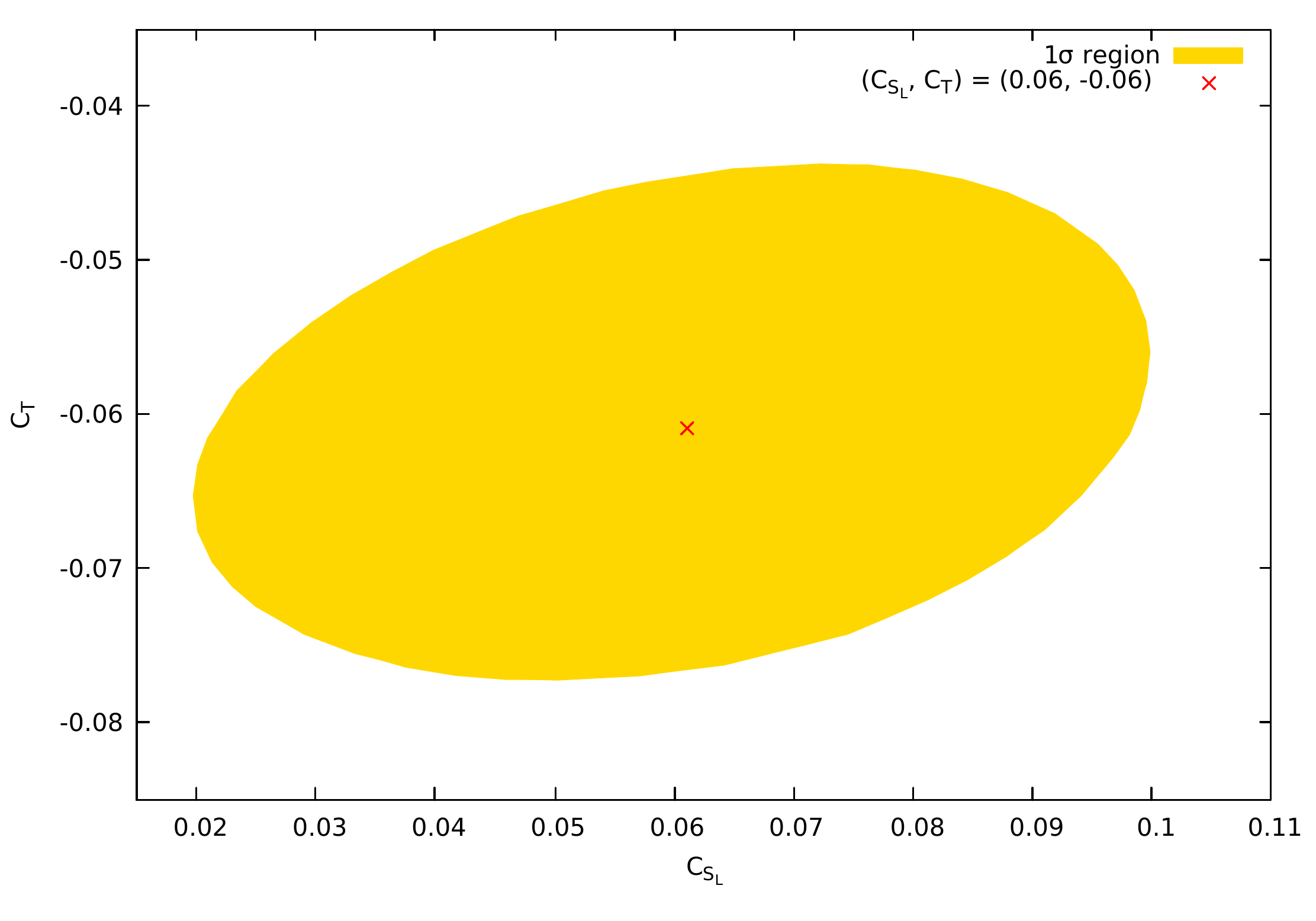}&
\includegraphics[width=65mm]{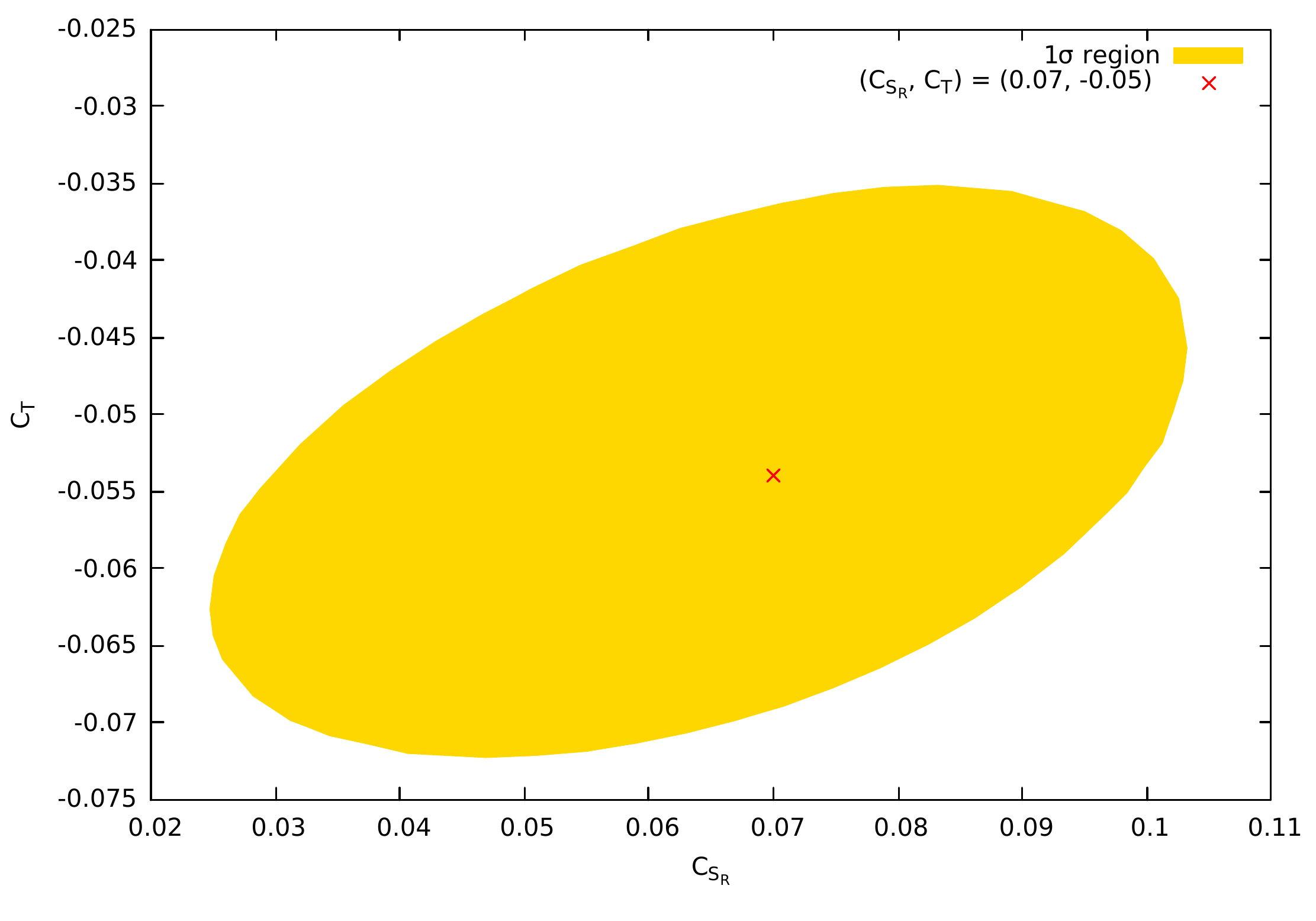}\\
\end{tabular}

\includegraphics[width=70mm]{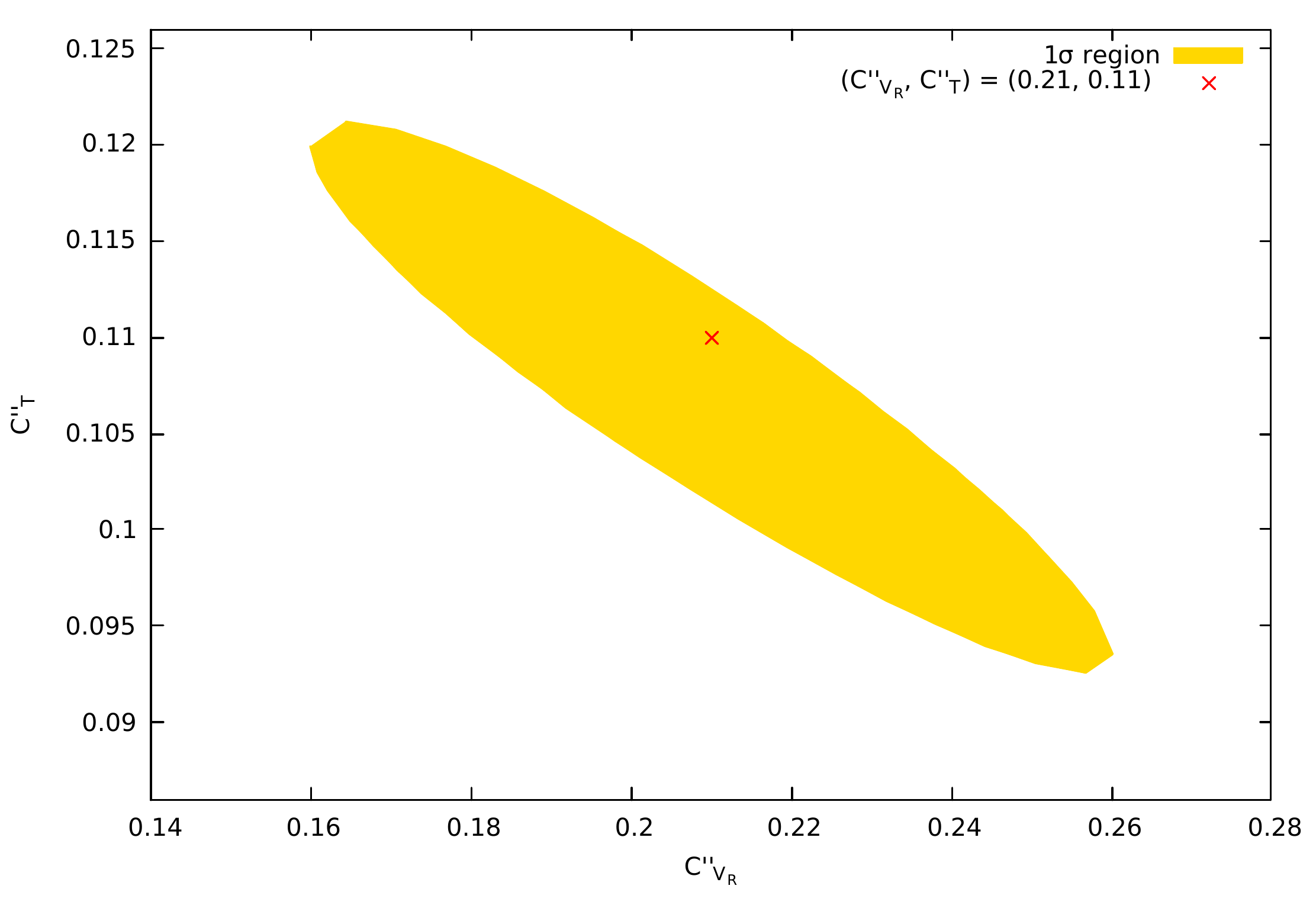}\\
\caption{The $1\sigma$ error ellipses for the two parameter solutions listed in table~\ref{two}. The best fit point is represented by red cross.}
\label{ellipses}
\end{figure*}

Table~\ref{two} lists best fit points of three solutions with two dis-similar operators, along with the $\mathcal{O}_{V_L}$ solution. As before, these solutions also satisfy $\chi^2_{\rm min}\leq 5$ as well as $\mathcal{B}(B_c\rightarrow \tau\bar{\nu})< 10\%$. The $1\sigma$ error ellipses for these solutions are shown in fig.~\ref{ellipses}. If a looser constraint $\chi^2_{\rm min}\leq 6.0$ is used, we obtain two additional solutions: $(C'_{V_R},\, C'_{S_L}) = (0.38, 0.63)$  and $(C''_{V_R},\, C''_{S_L}) = (0.11, -0.58)$.~\footnote{Recently it was claimed in ref.~\cite{Bardhan:2019ljo} that the present data allows a tensor solution with a small WC $C_T$. We find that a solution with $C_T = -0.07\pm 0.02$ occurs with $\chi^2_{\rm min}$ of $7.1$~\cite{Kumbhakar:2019avh}.}


 \begin{table}[htbp]
 \centering
 \tabcolsep 7pt
 \begin{tabular}{|c|c|c|}
\hline\hline
NP type & Best fit value(s)& $\chi^2_{\rm min}$  \\
\hline
SM  & $C_{i}=0$ & $21.80$  \\
\hline
$C_{V_L}$  &  $0.10 \pm 0.02$& $4.5$ \\


\hline
$(C_{S_L}, C_T)$ & $(0.06, -0.06)$ & 5.0 \\
\hline
$(C_{S_R}, C_T)$ & $(0.07, -0.05)$ & 4.6 \\
\hline
$(C''_{V_R}, C''_{T})$ & $(0.21, 0.11)$ & 4.2 \\
\hline\hline	



\end{tabular}
\caption{Best fit values of the coefficients of new physics operators at $\Lambda = 1$ TeV by making use of data of $R_D$, $R_{D^*}$, $R_{J/\psi}$, $P_{\tau}^{D^*}$ and $f_L^{D^*}$. In this fit, we use the HFLAV summer 2019 averages of $R_D$-$R_{D^*}$. Here we list the solutions for which $\chi^2_{\rm min}\leq 5.0$ as well as $\mathcal{B}(B_c\rightarrow \tau\,\bar{\nu})< 10\%$.}
\label{two}
 \end{table}

In table~\ref{three}, We have listed the predictions  for the five experimental observables which went into the fit for each of the allowed solutions. The set of predictions for each solution matches the measured values very well.   
\begin{table}[htbp]
\centering
\tabcolsep 7pt
\begin{tabular}{|c|c|c|c|c|c|}
\hline\hline
NP type & $R_D$ & $R_{D^*}$ & $R_{J/\psi}$ & $P_{\tau}^{D^*}$ & $f_L^{D^*}$\\
\hline
SM  &$0.297\pm 0.008$ &$0.253\pm 0.002$ & $0.289\pm 0.010$ &$-0.498\pm 0.004$ &  $0.46\pm 0.04$ \\
\hline
$C_{V_L}$  &$0.343\pm 0.010$ &$0.292\pm 0.005$ & $0.335\pm 0.012$ &$-0.499\pm 0.005$ &  $0.46\pm 0.04$\\

\hline
$(C_{S_L}, C_T)$ & $0.337\pm 0.011$ & $0.295\pm 0.003$ & $0.345\pm 0.009$ & $-0.481\pm 0.007$ & $0.44\pm 0.06$\\
\hline
$(C_{S_R}, C_T)$ & $0.345\pm 0.009$ & $0.292\pm 0.004$ & $0.341\pm 0.011$ & $-0.461\pm 0.007$ & $0.45\pm 0.04$\\
\hline
$(C''_{V_R}, C''_{T})$ & $0.349\pm 0.010$ & $0.300\pm 0.006$ & $0.353\pm 0.012$ & $-0.425\pm 0.010$ & $0.46\pm 0.05$\\
\hline\hline


\end{tabular}
\caption{The predictions of $R_D$, $R_{D^*}$, $R_{J/\psi}$, $P_{\tau}^{D^*}$ and $f_L^{D^*}$ for each of the allowed NP solutions.}
\label{three}
\end{table}


In order to discriminate between the four allowed solutions, we consider some of the other observables which can be measured in decays driven by the $b\rightarrow c\,\tau\,\bar{\nu}$ transition. In particular, we consider the following angular observables in $B\rightarrow (D,D^*) \,\tau \,\bar{\nu}$~\cite{Hu:2018veh,Murgui:2019czp,Shi:2019gxi,Blanke:2019qrx,Becirevic:2019tpx}:
\begin{itemize}
\item The $\tau$ polarization $P_{\tau}^D$ in $B\rightarrow D \,\tau \,\bar{\nu}$
\item The forward-backward asymmetry $A^D_{FB}$ in $B\rightarrow D \,\tau\, \bar{\nu}$
\item The forward-backward asymmetry $A^{D^*}_{FB}$ in $B\rightarrow D^*\, \tau\, \bar{\nu}$
\item The branching ratio of $B_c\rightarrow \tau\,\bar{\nu}$.

\end{itemize}
The predictions of each of these quantities for the four solutions are listed in table~\ref{four}.
\begin{table}
\centering
\tabcolsep 7pt
\begin{tabular}{|c|c|c|c|c|}
\hline\hline
NP type  & $P_{\tau}^D$ & $A^{D}_{FB}$& $A^{D^*}_{FB}$ &  $\mathcal{B}(B_c\rightarrow \tau\bar{\nu} )\, \%$ \\
\hline
SM & $0.324\pm 0.001$& $0.360\pm 0.001$ & $-0.012\pm 0.007$  & 2.2\\
\hline
$C_{V_L}$  & $0.324\pm 0.002$ & $0.360\pm 0.002$ &$-0.013\pm 0.007$  &2.5\\
\hline
$(C_{S_L}, C_T)$ & $0.442\pm 0.002$ & $0.331\pm 0.003$ & $-0.069\pm 0.009$ & 0.8 \\
\hline
$(C_{S_R}, C_T)$ & $0.450\pm 0.003$ & $0.331\pm 0.002$ & $-0.045\pm 0.007$  &4.0 \\
\hline
$(C''_{V_R}, C''_{T})$ & $0.448\pm 0.002$ & $-0.244\pm 0.003$ & $-0.025\pm 0.008$  & 11.0\\
\hline
\hline
\end{tabular}
\caption{The predictions of $P^D_{\tau}$, $A_{FB}^D$, $A^{D^*}_{FB}$ and $\mathcal{B}(B_c\rightarrow \tau\,\bar{\nu})$ for each of the allowed NP solutions.}
\label{four}
\end{table}
\begin{figure}
\centering
\includegraphics[width=90mm]{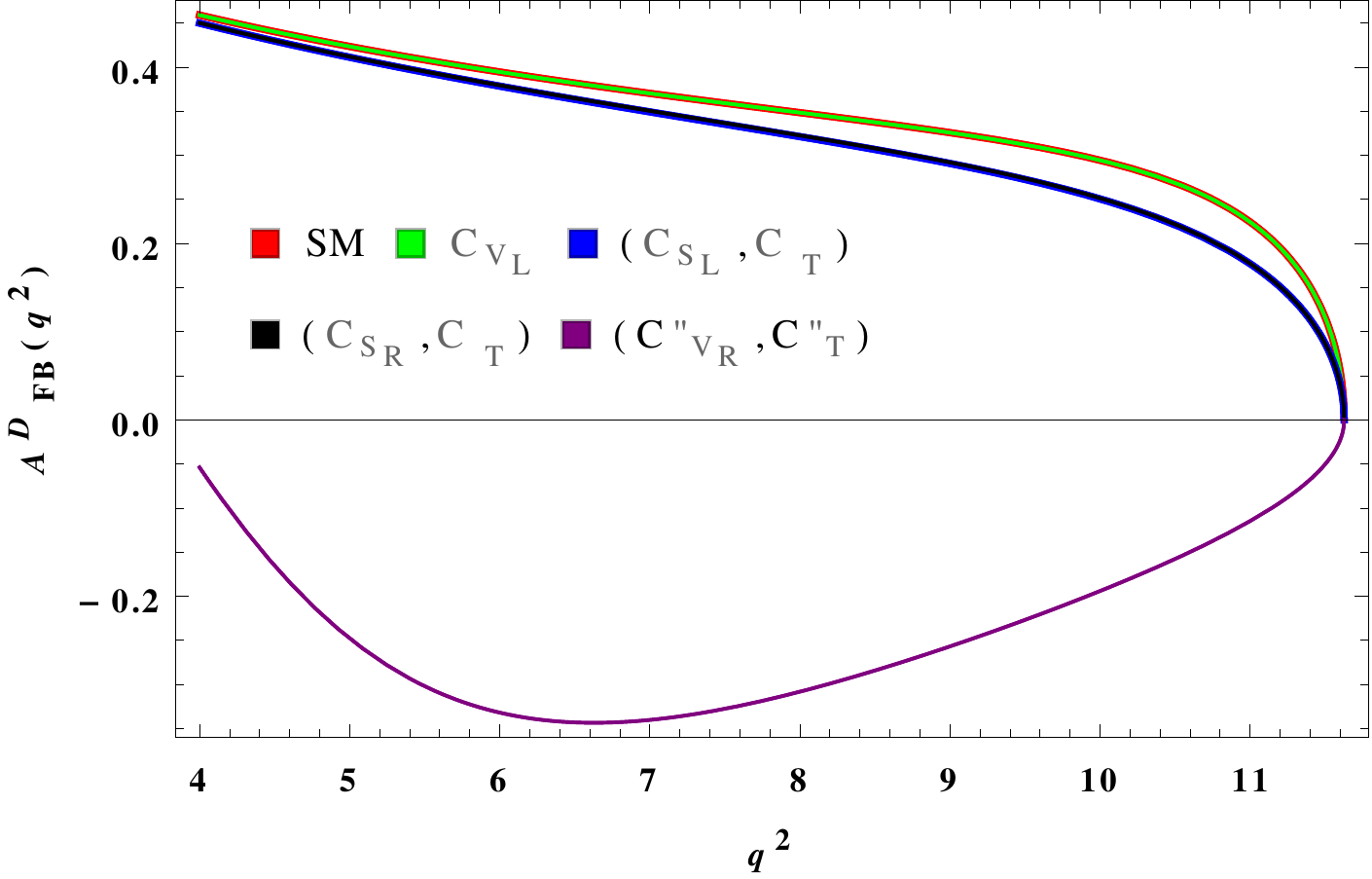} 
\caption{The plot corresponds to $A_{FB}^{D}(q^2)$ in $B\rightarrow D\tau\bar{\nu}$ decay. The bands in this figure represent $1\sigma$ range which is mainly due to various form factors and is obtained by adding these errors in quadrature. The color code for each NP solution as well as the SM  is shown in the figure.}
\label{fig2}
\end{figure}

From table~\ref{four} we note the following distinguishing features:
\begin{itemize}
\item We see that $P^D_{\tau}$ and $A^{D^*}_{FB}$ have poor distinguishing ability.

\item A measurement of $\mathcal{B}(B_c\rightarrow \tau\bar{\nu})$ to an accuracy of $2\%$ can make a distinction between all four solutions.

\item The asymmetry $A^D_{FB}$ can distinguish $(\mathcal{O}''_{V_R},\, \mathcal{O}''_{T})$ solution from the other four. This is also illustrated in fig.~\ref{fig2}.


\end{itemize}

\section{Conclusions}

The new measurements of $R_D$ and $R_{D^*}$, announced by the Belle Collaboration at Moriond 2019, reduced the discrepancy between the 
SM predictions and the global average values from $4.1~\sigma$ to $3.1~\sigma$. The measured value of $f_L^{D^*}$ very strongly
discriminates against tensor NP solutions with large WC. In this work, we did a fit with the new global averages and found that there are
only {\bf four} allowed NP solutions. We also explored the possibility of making a distinction between these solutions by measuring various 
angular asymmetries in $B \to D/D^* \,\tau\, \bar{\nu}$, $\tau$ polarization asymmetry in $B \to D\, \tau \,\bar{\nu}$ and the branching
ratio $\mathcal{B}(B_c\rightarrow \tau\,\bar{\nu}$). We found that each of these four solutions can be uniquely identified by
the measurement of the branching ratio of $B_c\rightarrow \tau\,\bar{\nu}$ to a precision of $2\%$. 
\section*{Acknowledgement}
The work of DK is partially supported by the National Science Centre (Poland) under the research grant No.2017/26/E/ST2/00470. 


\end{document}